# A State of Art Review on Wireless Power Transmission Approaches for Implantable Medical Devices

Mohammad Haerinia
*School of Electrical Engineering and Computer Science, University of North Dakota,
Grand Forks, ND 58202, USA
Email: Mohammad.haerinia@und.edu*

Wireless power transmission (WPT) is a critical technology that provides a secure alternative mechanism for wireless power and communication with implantable medical devices. WPT approaches for implantable medical devices have been utilized based on applications. For instance, the inductive coupling tactic is mostly employed for transmission of energy to neuro-stimulators, and the ultrasonic method is used for deep-seated implants. This article provides a study concentrating on popular WPT techniques for implantable medical devices (IMDs) including inductive coupling, microwave, ultrasound, and hybrid WPT systems consisting of two approaches combined. Moreover, an overview of the major works is analyzed with a comparison of their major design elements, operating frequency, distance, efficiency, and harvested power.

## 1. INTRODUCTION

In recent years, medical progress has evolved with an increased interest in instruments for sensing and controlling the specific functions of the brain. These medical instruments considerably decrease morbidity and improve the standard of life for certain patients. Sensor systems are now quite advanced but providing power to these devices is still a major challenge. The answer to this issue is using WPT technologies for a range of biomedical implants. WPT is a secure and appropriate energy supply for recharging biosensors and electrical implanted devices as well as for data communication in these specific applications. The biomedical implants are intended to be used for biological studies, therapy, and medical diagnostics. Not to mention that novel biological materials provide additional biocompatibility, efficiency as well as reduced expenses. Implantable medical devices (IMDs) can be classified into two primary categories based on their methodologies for the transmission of power. Transfer mechanisms such as inductive coupling, optical charging, and ultrasound are included in the first category. The second category is split into two subsections. Batteries such as lithium and natural harvesting, including biofuel cell, thermoelectricity, piezoelectricity, electrostatic, electromagnetic [1]. Various WPT techniques are reviewed in the literature. For instance, the ultrasound and inductive coupling methods were evaluated by Taalla et al. [2] and Shadid et al. [3], respectively. In this paper, common WPT approaches for IMDs including, inductive coupling, microwave, and ultrasound are studied. Hybrid WPT systems, a mixture of two different methods, are also reviewed.

## 2. DIFFERENT APPROACHES FOR A WIRELESS POWER TRANSFER SYSTEM

The lifespan of implantable medical devices is limited to battery capabilities. Patient pain and the danger of infection are the major development concerns in implantable medical systems because the use of implanted batteries can cause diseases [4]. Therefore, the WPT link is a safer option to power biomedical implants [4]. Typically, non-rechargeable batteries with greater weight and volume, and shorter period of effectiveness compared to rechargeable batteries, are employed for implantable medical devices. Medical implants like implanted spinal cord stimulators can use a rechargeable battery to improve their capability and reduce overall costs [5]. Lately, there is a great interest in the usage of the WPT for medical applications. The development of implantable electronic devices in a biological system made it easier to use this technology for powering various implantable medical devices such as biological sensors, pacemakers, and neurostimulator working in a range of power from a few microwatts to a few watts. As shown in Figure 1, the power ranges of common implantable medical devices are illustrated [1], [6]. The WPT system for the neurostimulators and the pacemaker are discussed in detail in [7][8][9][10][11][12] and [13], respectively.

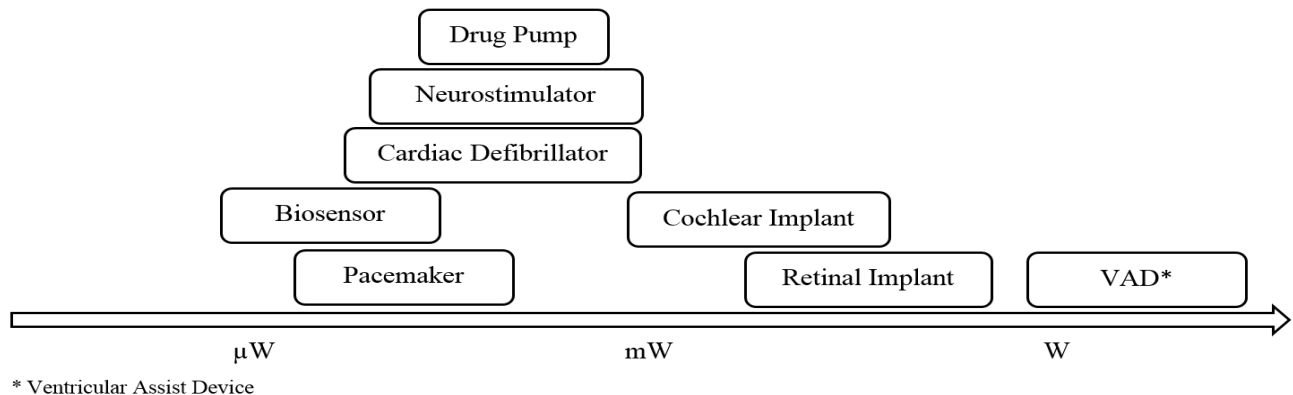

Figure 1: Power ranges of implantable medical devices.

There are reliability problems with the classic wireless power links. An option that facilitates the growth of bio-implants is the use of CMOS processes. The procedure is to include the standard CMOS with the implanted receiver. This reduces the cost, improves productivity, and provides compactibility and reliability of prototypes [14][15]. The usage of CMOS for WPT systems are presented in [16][17][18][19][15][20][21][22][23]. A backward communication unit transmits the information to an external data communicator using modulation. Typically FSK[24],PSK[25], ASK[26], OOK[27], LSK[28], PPSK[29],QMPM[30], QPSK[31], impedance modulation [32] have been used for data communication units in medical applications. Long-term RF and microwave exposure is dangerous. The device layout must, comply with the associated safety regulations to protect patients from electromagnetic radiation damage. It is possible to evaluate maximum permissible exposure (MPE) in environments for electromagnetic field intensity by assessing the specific absorption rate (SAR). IEEE Standard Basis C95.1 expresses SAR limitation. According to this principle, the maximum SAR value must be below 1.6 W/kg for any 1g of the body tissue and below 0.08 W/kg for the whole body. Nevertheless, the maximum SAR limitation is 4 W/kg for every 10 g of the tissue of the body parts such as the hands, feet, ankles, and wrists. Mainly, determining SAR can be achieved by using numerical techniques and empirical models using fabricated tissue phantoms [4] [33] [34] [35] [36][37][ 38].The maximum SAR value is studied in [39] [40]. The empirical results can be obtained, in vivo [41][42][8], using a living organism or in vitro [43][44][45], outside of a living organism. To mimic the biological effects of human body tissue, the phantom is very popular among researchers in this field. The tissue electromagnetic properties play an important role in the design of implantable devices. The assessment of variation on tissue electromagnetic properties was provided by Bocan et al [47]. The recent reports on tissue electromagnetic properties are depicted in Table.1.

Table I: Summary of different approaches in analyzing of tissue electromagnetic properties.

| Reference | Year | Tissue | Frequencies | Models/Methods |
|---|---|---|---|---|
| 48 | 2020 | in-vivo, ex-vivo | - | FEM* |
| 49 | 2019 | Muscle, fat, skin | 50MHz, 300MHz, 700MHz and 900MHz | FDTD** |
| 50 | 2019 | Body | (0.5-26.5) GHz | Measured properties, Cole-Cole |
| 51 | 2018 | Brain, liver | 200–1600 Hz | Measured properties |
| 52 | 2018 | Muscle, fat, skin | 915 MHz and 2 GHz | Measured properties |
| 53 | 2017 | Blood, liver, fat, and brain | 10 kHz-10MHz | Bottcher-Bordewijk model, Measured properties |
| 54 | 2016 | Muscle, bladder, cervix | 128 MHz | Measured properties, Cole-Cole |
| 55 | 2016 | Body/14 tissues | 2.1 GHz, 2.6 GHz | FDTD |
| 56 | 2016 | Head | (0.75 – 2.55) GHz | Phantom/ FEM |
| 57 | 2016 | Muscle | 500 MHz- 20 GHz | Fricke |
| 58 | 2015 | Eye/6 tissues | (0.9 – 10) GHz | FDTD |
| 59 | 2015 | Skin | (0.8 – 1.2) THz | FEM |
| 60 | 2014 | Eye, head/14 tissues | (0.9 – 5.8) GHz | FDTD |
| 61 | 2010 | Head | - | FEM |
| 62 | 2009 | Head/16 tissues | 50 MHz- 20 GHz | Measured properties, FDTD |
| 63 | 2006 | Eye, head/15 tissues | 900 MHz, 1800 MHz, 2450 MHz | FDTD |
| 64 | 2004 | Body | 400 MHz, 900 MHz, 2400 MHz | Visible human, FDTD |
| 65 | 2004 | Body /51 tissues | 30 MHz- 3 GHz | FDTD |
| 66 | 2002 | Head/10 tissues | 900 MHz, 1800 MHz | Visible human, FDTD |

* Finite Element Method
** Finite-Difference Time-Domain

## 2.1. Inductive-Based Wireless Power Transfer

Inductive coupling is a common and efficient way to transfer data and power into implantable medical instruments, including cardiac pacemakers, implantable cardioverter defibrillators, recording devices, neuromuscular stimulators, cochlear and retinal implants. The development of an inductive link with a power amplifier is applied-based and can be adjusted with operating frequency, range, and form factor and output power. The bandwidth to support data communication and reasonable efficacy for power transfer, insensitivity to misalignments, and biocompatibility are needed for a robust inductive link for medical implants [67]. In general, an inductive-based wireless power transfer system for IMDs has a pair of coils to inductively couple the power from a primary coil outside the body to a secondary implanted coil. Hundreds of kilohertz to a few megahertz is the operating frequency, and the size of the implanted coil is between several millimeters to a few centimeters. As the frequency increases, the electromagnetic wavelength gets more commensurate with the coil dimension and the space between the coils. In this stance, the radiative and non-radiative components are part of the electromagnetic waves. Biological tissue also creates significant problems for the propagation of electromagnetic fields and dilutes the electrical field, thus affecting the efficiency of the inductive link [33]. According to Faraday's induction law, increasing the size of coils and the number of turns boosts inductive link efficiency [33]. In case that the transmitting coil and the receiving coil have the same size, the maximum coupling is achievable. Although, in practical, the implanted coil is significantly smaller than the transmitting coil [68]. Mainly, the inductive-based wireless power transfer system is used for medical devices such as the brain and spinal cord stimulators. LYU et al. [8] have developed a stimulator, which occupies a dimension of 5 mm × 7.5 mm and operating at the resonant frequency of 198 MHz while having a 14 cm distance from the transmitter which is located outside of the body. The stimulator gets the energy that has already been stored by a switched capacitor and releases the energy as an output stimulus once the voltage reaches a threshold. The control unit utilizes positive feedback to trigger the circuit, so no stimulation control circuit block is needed. An in vivo experiment was performed to demonstrate the performance of the stimulator. Two electromyography (EMG) recording electrodes were implanted into the gastrocnemius muscle of a rat while the ground electrode attached on the skin. The proposed stimulator, which is implanted on the sciatic nerve of the rat and driven by a source located outside the body, is shown in Figure 2. A free-floating neural implant, which is insensitive to the location, is provided as an inductive link in [10] for wireless energy transmission. The authors have created prototypes of floating implants for precise measurements as shown in Figure 3.

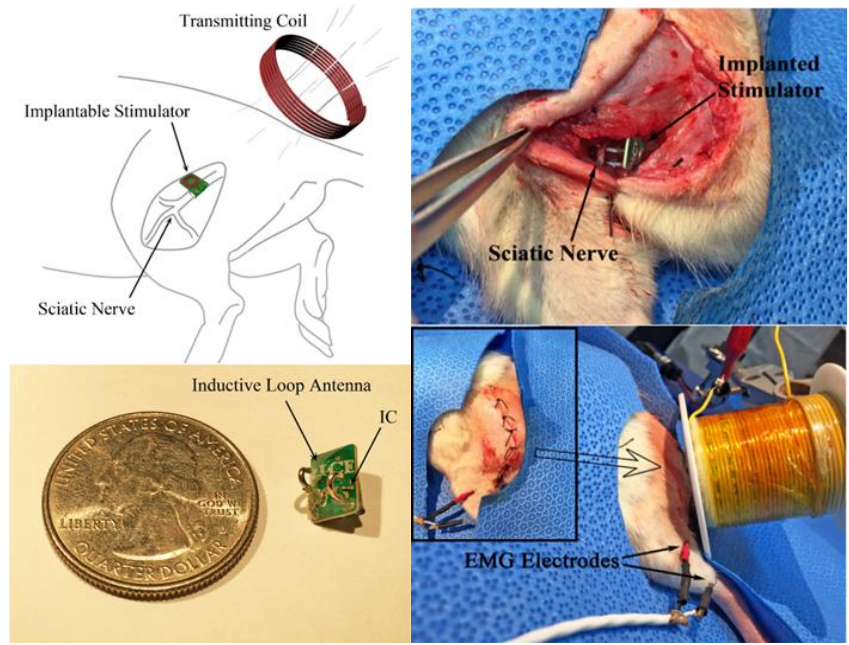

Figure 2: In vivo experiment with the stimulator fully implanted on the sciatic nerve [8].

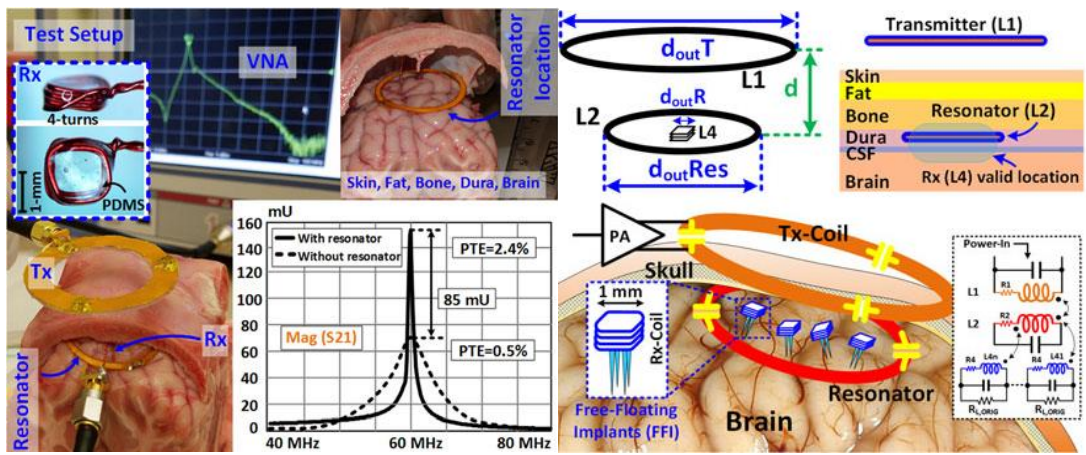

Figure 3: The WPT link for distributed free-floating implants encompassed by an implanted high-Q resonator [10].

The system works with a power transfer efficiency of 2.4 % at 60 MHz and provides 1.3 mW power to the implant 14–18 mm away from the transmitter. Their coil link is stable against the lateral and angular misalignments of the floating implants as long as the coils continue to have the high-Q resonator. The extra heat produced by the resonator coil also does not exceed safety limits. Recent works of the inductive WPT scheme are evaluated and presented in Table.2. The panel consists of printed and 3D coils. Printed coils maintain acceptable performance under lateral malalignment and are reliable for implants [4].

Table II: Existing inductive-based WPT approaches for implantable power applications.

| Reference | Year | Frequency | Output Power (mW) | Efficiency (%) | Active Range (mm) | Transmitter Dimension (mm) | Receiver Dimension (mm) |
|---|---|---|---|---|---|---|---|
| 69 | 2020 | 915 MHz | - | 1.93 | 40~50 | - | $30 \times 30$ |
| 70 | 2020 | 5.8 GHz | 0.01 | $1.2 \times 10^{-5}$ | 1 | - | 0.116×0.116 |
| 71 | 2019 | 13.56 MHz | - | 79.6 | 15 | $66 \times 70$ | $66 \times 70$ |
| 7 | 2019 | 430 MHz | 0.0033 | - | 45 | - | $4.5 \times 3.6$ |
| 33 | 2019 | 434 MHz | 31.62 | 0.68 | 10 | $20 \times 20$ | $1.6 \times 1.6$ |
| 8 | 2018 | 198 MHz | >0.0027 | - | 140 | $d_{outT} = 30.5$ | $d_{outR} = 4.9$ $d_{inR} = 2.2$ |
| 40 | 2018 | 60, 300, MHz330 | - | 2.12, 3.88, 1.68 | 12 | $d_{outT}$ =17.2, 24, 26 | $d_{outR}$ =4 |
| 72 | 2018 | 2, 4 MHz | 126 | 25 | 6 | $d_{outT} = 35$ | $d_{outR} = 20$ |
| 9 | 2018 | 1.3GHz | 3981 | - | 5 | $d_{outT}$ =10 | $d_{outR}$ =0.2 |
| 34 | 2018 | 39.86 MHz | 115 | 47.2 | - | $d_{outT} = 63.9$ $d_{inT} = 34.08$ | $d_{outR} = 21.56$ $d_{inR} = 11.52$ |
| 73 | 2018 | 432.5MHz | 1.05 | 13.9 | 10 | - | - |
| 74 | 2018 | 430 MHz | - | - | 60 | $30 \times 30$ | $10 \times 10$ |
| 75,76 | 2018 | 3 MHz | 772.8 | 38.79 | 5-15 | $d_{outT}$ =45.2 $d_{inT} = 10$ | $d_{outR}$ =36.4 $d_{inR} = 10$ |
| 77 | 2018 | 2.75 MHz | - | - | - | $d_{outT}$ =45.2 $d_{inT} = 10$ | - |
| 11 | 2017 | 13.56 MHz | 18 | 7.7, 11.7 | 10 | $d_{outT} \approx 30$ | $d_{outR} = 10$ $d_{inR} = 4.6$ |
| 39 | 2016 | 50MHz | 0.0657 | 0.13 | 10 | $d_{outT}$=21 | $d_{outR}$=1 |
| 19 | 2014 | 8.1MHz | 29.8~93.3 | 47.6~65.4 | 12~20 | $d_{outT}$ =30 | $d_{outR}$ =20 |
| 21 | 2019 | 12.85 MHz | - | 75.8 | - | $30.0 \times 29.6$ | $30.0 \times 29.6$ |
| 78 | 2019 | 1-100 MHz | - | - | 15 | - | $d_{outR} = 1.75$ $d_{inR} = 0.50$ |
| 41 | 2018 | 433 MHz | 0.0008 | 0.87 | 600 | - | $d_{outR}$ =10 $d_{inR}$ =9.2 |
| 29 | 2017 | 13.56MHz | ≤100 | - | 5-15 | $d_{outT}$=25 | $d_{outR}$=16 |
| 10 | 2017 | 60 MHz | 1.3 | 2.4 | 16 | $d_{outT} = 45$ $d_{inT} = 35$ | $d_{outR} = 1.2$ $d_{inR} = 1$ |
| 36 | 2016 | 20MHz | 2.2 | 1.4 | 10 | $d_{outT}$=20,28 | $d_{outR}$=1 |
| 42 | 2016 | 40MHz | - | 2.56 | 70 | $d_{outT}$=100 | $d_{outR}$=18 |
| 79 | 2016 | 10 kHz-50 kHz | - | - | - | $d_{inT}$ =4.5 | - |
| 30 | 2015 | 2MHz | 1450 | 27 | 80 | $d_{outT}$=140 | $d_{outR}$=65 |

## 2.2. Microwave-Based Wireless Power Transfer

Another way to efficiently transmit power wirelessly over long distances in the order of meters to kilometers is the microwave power transmission. The challenges for this technique in present day include the minimization of energy loss, protecting both humans and animals against exposure to excessive microwave radiation, and reconfiguring of a wireless transmission system in reaction to modifications such as a shift in a range between transmitter and receiver [80]. Pacemaker implantation is a popular method to cure people with cardiac insufficiency. Although, the lifetime of the pacemaker is restricted to the lifespan of the battery, and the installation of a subcutaneous pocket [13]. Asif et al. [13] built a rectenna-based leadless pacemaker prototype. For energy transmission to the implanted unit, a wearable transmitting antenna range was fabricated. To evaluate the system's efficiency through Vivo ECG outcomes, an animal study is implemented as shown in Figure 4.

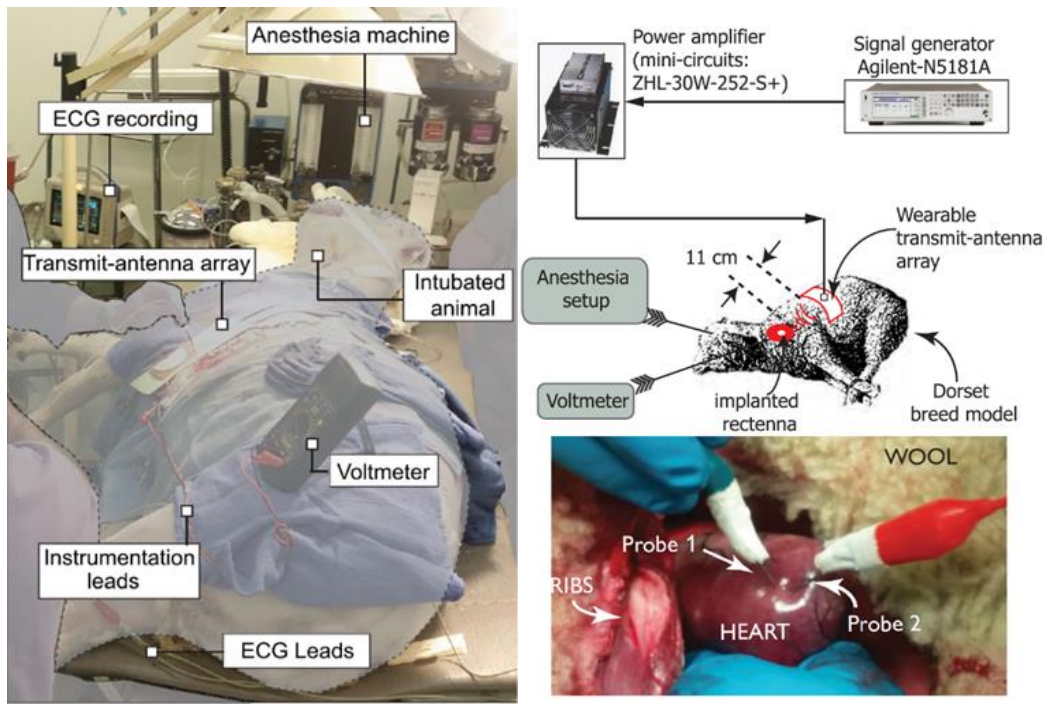

Figure 4: A complete layout of the in vivo experiment setup used to examine the proposed rectenna-based leadless pacemaker [13].

The authors assert that the calculations of specific absorption rate (SAR) are within the limits suggested by IEEE and also claim that the proposed leadless pacing method is safer as well as eliminating the battery, leads, and device pocket. Zada et al [81] provided a miniaturized implantable antenna having three frequency bands (902-928 MHz, 2400-2483.5MHz, and 1824–1980 MHz) operating at the industrial, scientific, and medical (ISM) band and midfield band, respectively. A capsule-shaped and a flat type antenna are fabricated with a volume of 647 $mm^3$ and 425.6 $mm^3$, respectively. Figure 5 shows a full design of devices and measurement setup. This triple band antenna is complemented with microelectronics, sensors, and batteries for stimulation in different applications. The system is examined in different tissues including the scalp, heart, colon, large intestine, and stomach.

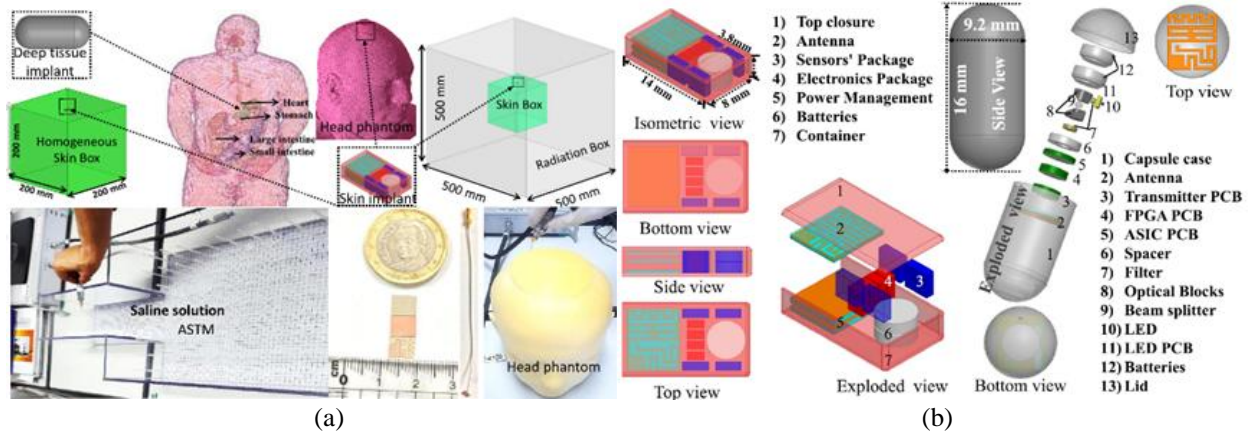

Figure 5: Analysis setup of the proposed antenna. (a) Simulation and measurement setups (b) Comprehensive device architecture [81].

Asif et al [82] took advantage of a microwave-based wireless power transfer technique to charge deep medical implants like cardiac pacemakers. Their novel wideband numerical model (WBNM) was to provide an RF power source of a leadless pacemaker while using a metamaterial-based antenna operating at 2.4 GHz. They used the tissue simulating liquid (TSL) mimicking the human body to prove the performance of their design for implantable applications. A wireless powering technique was introduced by Ho et al [83], which overcomes the difficulty of miniaturizing the power source via adaptive electromagnetic energy transport. This method is designed for micro-implants like microelectromechanical systems sensors and optoelements. Recent works of microwave-based WPT system are reviewed and shown in Table.3.

Table III: Existing microwave-based WPT approaches for implantable power applications.

| Reference | Year | Frequency | Output Power (mW) | Efficiency (%) | Active Range (mm) | Transmitter Dimension (mm) | Receiver Dimension (mm) |
|---|---|---|---|---|---|---|---|
| 84 | 2020 | 1.47GHz | 6.7 | 0.67 | 50 | 6 × 6 | - |
| 85 | 2020 | 0.403GHz, 2.44GHz | - | - | 30-350 | - | 9.5 × 9.5 |
| 86,87, 88 | 2019 | 1.64GHz, 3.56GHz | - | 32,1.1 | - | 14 × 15 | 14 × 15 |
| 13 | 2019 | 954 MHz | 10 | 65 | 110 | - | 10 × 12 |
| 81 | 2018 | 0.915, 1.9, 2.45 GHz | 0.398 | - | 4.5 | - | 7 × 6 |
| 37 | 2018 | 400MHz | 19,82 | - | 1,3,6,12,15 | $d_{outT}$ =18 $d_{inT}$ = 12 | 1 × 1 |
| 38 | 2018 | 13.9 MHz | 100 | 0.39 | 50 | 0.3 × 1.3 | 8 × 28 |
| 89 | 2018 | 280 MHz | 44 | - | 3 | 30 × 80 | - |
| 112 | 2017 | 2.45 GHz | 2280,600, 240,96 | - | 1000-4000 | - | $d_{outR}$ =63.6 |
| 144 | 2014 | 2.4 GHz | - | 15-78 | 10-100 | 63×39×50 | 63×39×50 |

## 2.3. Ultrasonic-Based Wireless Power Transfer

The ultrasound imaging is a well-known tool for evaluating patients' physiological and pathological conditions. In the passive ultrasonic recorder, the backscattered echo is derived from the reaction of biological tissue's acoustic properties to sound waves. Besides, the acoustic emission can be used for supplying energy wirelessly in the active biological environment [44]. The ultrasonic-based wireless power transfer system has a transmitter converting electrical energy to ultrasonic energy, and also a receiver converting back the ultrasonic energy to the electrical energy. The ultrasonic-based WPT system is an effective way for medical applications such as a cardiac defibrillator and a deep brain stimulator [90]. A mode of clinical therapy is a stimulation of excitable tissue for different disorders such as Parkinson's disease, urinary incontinence, heart arrhythmia. The traditional stimulus techniques use percutaneous cables to transport electricity to the electrodes. The classical techniques are dangerous because they can cause infection [12]. The ultrasound or inductive based wireless power transfer is an interesting solution for this application. The advantage of ultrasound compared to magnetic resonance and induction coupling is that those methods are restricted to a short transfer distance, the misalignment issues [91], and magnetic field intensity should be under specified limitation for the safety of body exposure. In the ultrasonic method, the operating frequency needs to be changed according to sound radiation and pressure distribution to obtain the optimum energy transition situation [90]. In the range of frequencies individuals hear, Kim et al. [91] have developed an implantable pressure sensing system driven by mechanical vibration. The pressure inductor has a planar coil with a center of ferrite in which their distance differs from the involved stress. An implantable pressure sensor prototype is designed, as shown in Figure 6, and examined in vitro and Vivo. The acoustic receiver is a piezoelectric cantilever and charges a capacitor by converting sound vibration harmonics to electrical energy. The stored electric charge will be discharged across an LC tank with an inductor sensitive to pressure during the period that the cantilever is not shaking.

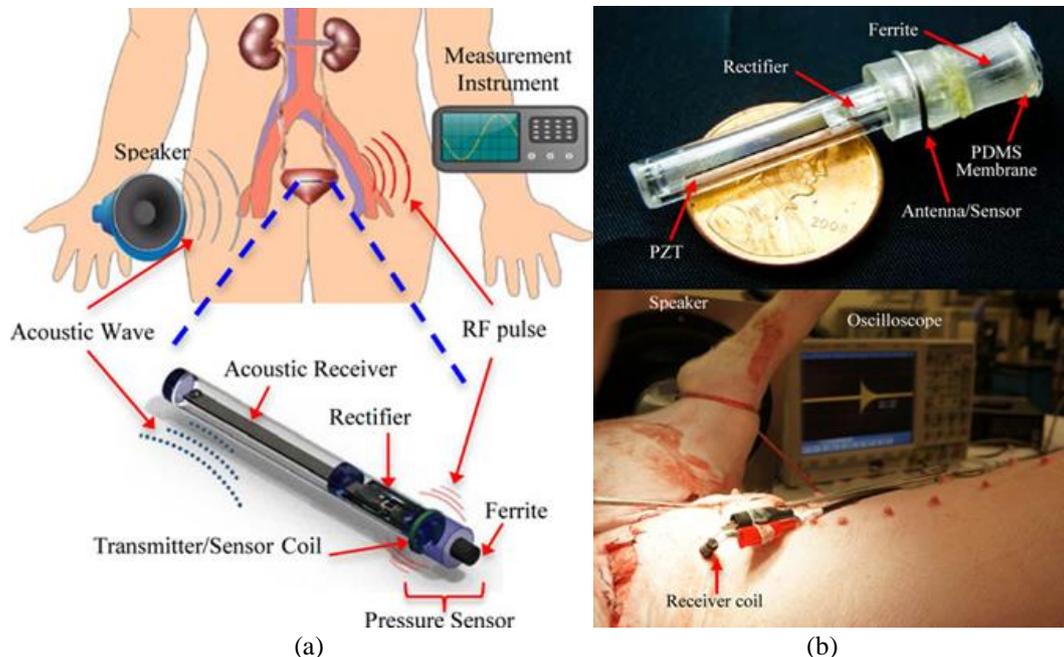

Figure 6: Proposed implantable pressure sensing system (a) schematic of an acoustically powered LC transponder implanted in the bladder (b) in vivo experiment setup [91].

Song et al. [92] investigated omnidirectional ultrasonic powering for deep implantable microdevices. During testing the omnidirectionality and outcome of the power transmission under the acoustic FDA regulations, the piezoelectric devices with distinct geometries were examined. The receivers can produce power in a range of milliwatts with a matched load located 200 mm far from them. The receivers have a symmetric geometry of $2 \times 2 \times 2$ $mm^3$ and are insensitive to misalignment. Figure 7 shows the proposed omnidirectional ultrasonic power system. Recent works of ultrasonic-based WPT system are shown in Table.4.

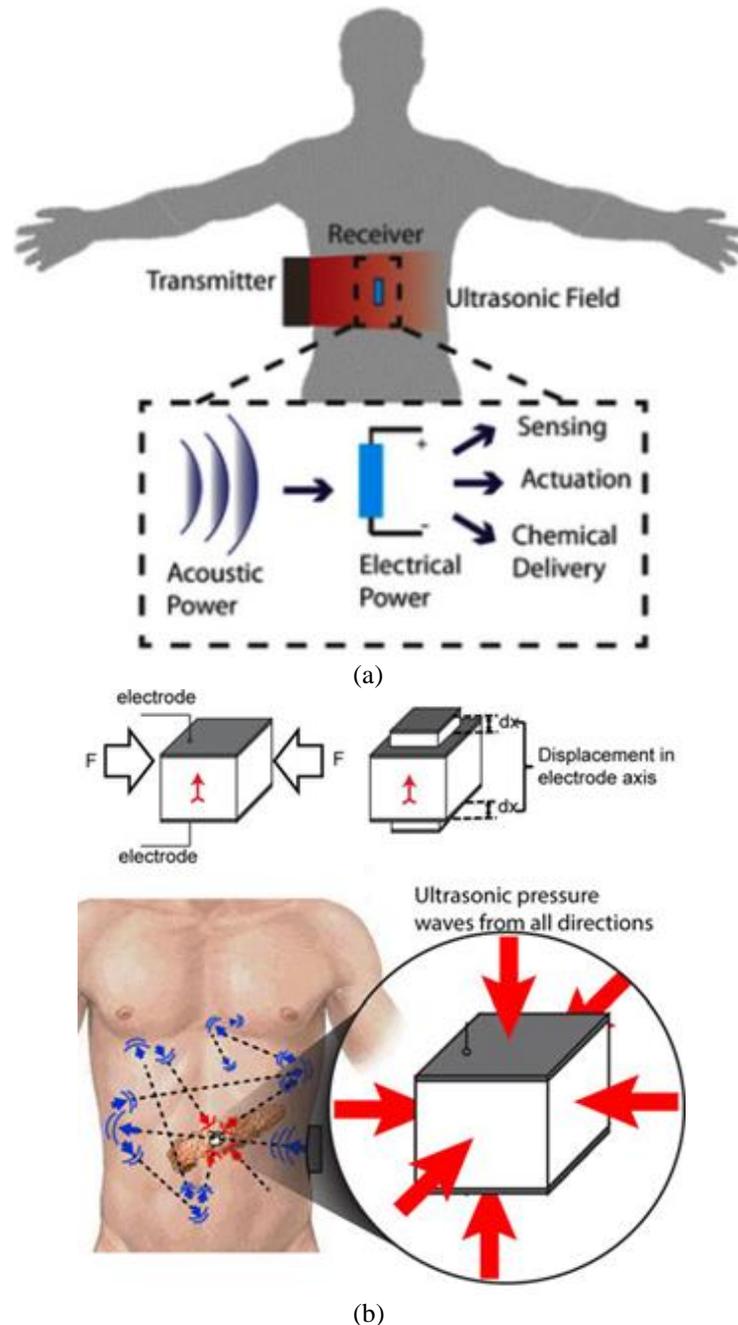

(a)

(b)

Figure 7: Proposed implantable omnidirectional ultrasonic powering system (a) schematic showing ultrasonic powering for an implantable medical device (b) factors contributing to omnidirectionality [92].

Table IV: Existing ultrasonic-based WPT approaches for implantable power applications.

| Reference | Year | Frequency | Output Power (mW) | Efficiency (%) | Active Range (mm) | Transmitter Dimension (mm) | Receiver Dimension (mm) |
|---|---|---|---|---|---|---|---|
| 93 | 2020 | 2.4 GHz | - | - | 200 | - | $d_{outR}$ = 10 |
| 94 | 2017 | 1MHz | 0.1 | - | 85 | $d_{outT}$ = 0.55 | - |
| 95 | 2017 | 1.8MHz | - | 2.11 | 30 | $d_{outT}$ = 10.8,15.9 | $d_{outR}$ = 1.1,1.2 |
| 16 | 2016 | 1MHz | 0.184 | - | - | - | - |
| 44 | 2016 | 400kHz | - | - | 30 | - | - |
| 96 | 2016 | 1MHz | - | 25 | 3-7 | $d_{outT}$ = 8 | - |
| 97 | 2015 | 280kHz | 2.6 | 18 | 18 | $d_{outT}$ = 20 | $d_{outR}$ = 20 |
| 98 | 2015 | 3.4MHz | 0.001 | - | 100 | - | - |
| 17 | 2015 | 30MHz | 0.1 | - | <100 | - | $d_{outR}$ =0.7,1 |
| 94 | 2014 | >350kHz | 0.016 | - | 150 | $d_{outT}$ = 8 | - |
| 18 | 2014 | 1MHz | 28 | 1.6 | 105 | $d_{outT}$ = 29.6 | $d_{outR}$ = 1 |
| 99 | 2013 | 1.07MHz | - | 45 | - | - | - |
| 94 | 2013 | 250kHz | 2.6 | 21 | 23 | $d_{outT}$ = 50 | $d_{outR}$ = 50 |
| 100 | 2011 | 1.2MHz | 100 | 50 | - | $d_{outT}$=44 | - |
| 45 | 2011 | 2.3MHz | ≈0.3 | - | 30-400 | $d_{outT}$=8 | - |
| 12 | 2011 | 1MHz | 23 | - | 120 | - | $d_{outR}$=8 |
| 101 | 2010 | 35kHz | 1.23 | - | 70 | - | $d_{outR}$=7 |
| 102 | 2010 | 673kHz | 1000 | 27 | 40 | - | - |
| 103 | 2003 | 100kHz | 5400 | 36 | 40 | - | - |
| 25 | 2002 | 1MHz | 2100 | 20 | 40 | - | - |
| 26 | 2001 | 1MHz | - | 20 | 30 | - | - |

## 2.4. Hybrid-Based Wireless Power Transfer

There is a hybrid wireless power transfer system in such a way a combination of two common methods working as a unit system. The advantage of a hybrid system is occupying less space than separate units. In this case, there is a capability to have different operating bands, more alternatives, and back-up. A hybrid inductive-based and microwave-based WPT system are presented in [104]. One of the current challenges for wireless transfer for small sensors is to minimize the system size. Haerinia, et al. [104] decreased the size of the compact system, at the same time implementing multi-functionality. This goal is obtained by designing an antenna having 14 mm×15 mm dimension and 20 mm×20 mm dimension for the hybrid system including antenna and coil. The coils operating frequencies are 510 MHz and the antennas work at 2.48 GHz and 4.66 GHz. Meng et al. [105] have developed a hybrid inductive-ultrasonic WPT link to power biomedical implants over bone, air, and tissue. They optimized cascaded inductive and ultrasonic links for WPT applications. The hybrid link is designed for an air–tissue medium to operate at 1.1 MHz with a power transmission efficiency of 0.16%. An ultrasonic transducer with a volume of 1 $mm^3$ is located 3 cm inside castor oil as shown in Figure 8. Recent works related to the hybrid WPT system are shown in Table.5.

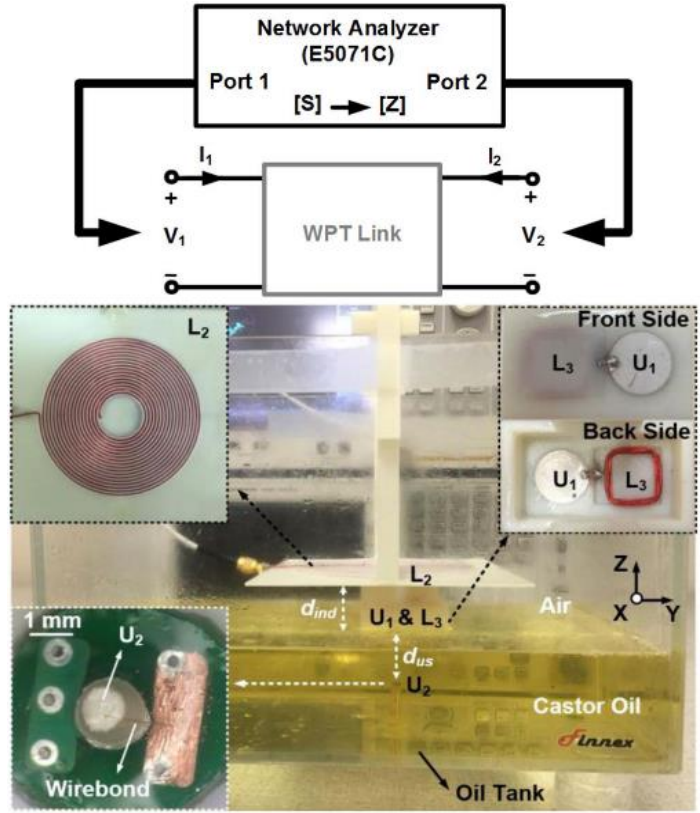

Figure 8: The power transfer efficiency measurement setup using a network analyzer [105].

Table V: Existing hybrid WPT approaches for implantable power applications.

| Reference | Year | Frequency | Output Power (mW) | Efficiency (%) | Active Range (mm) | Transmitter Dimension (mm) | Receiver Dimension (mm) | Methods |
|---|---|---|---|---|---|---|---|---|
| 104 | 2019 | 510 MHz, 2.48 GHz, 4.66 GHz | 0.0004 | 3.7, 2.2, 1 | 20-60 | $d_{outT}$ =39.75 $d_{inT}$ =16.25 20 ×20 | $d_{outR}$ = 20 $d_{inR}$ =15.95 14 ×15 | Inductive and Microwave |
| 106, | 2018 | 4 MHz | 500,53,53 | 1.9,2.6,0.98 | >30,15,30 | - | $d_{outR}$=40,83,83 | Inductive and Capacitive |
| 107,46 | 2018,2020 | 13.56 MHz, 415 MHz, 905 MHz, 1300 MHz | - | 10,0.5, 4.6,6.5 | 15-110 | $d_{outT}$ = 79.6 $d_{inT}$ = 54 | $d_{outR}$ = 31.5 $d_{inR}$ = 23 | Inductive and Microwave |
| 108 | 2017 | 13.56 MHz /910MHz | - | 17 | 16 | $d_{outT}$ = 83.2 $d_{inT}$ = 59.6 | $d_{outR}$ = 24.2 $d_{inR}$ = 19.2 | Inductive and Microwave |
| 105 | 2017 | 1.1 MHz | - | 0.16 | 60 | $d_{outT}$ = 100 | $d_{outR}$ = 15 | Inductive and Ultrasonic |
| 109 | 2012 | 200kHz | 8 | 1 | 70 | $d_{outT}$ = 39 $d_{inT}$ = 13 | $d_{outR}$ = 33 $d_{inR}$ = 11 | Inductive and Ultrasonic |

## 3. CONSIDERATION FOR DESIGN OF MEDICAL IMPLANTS AND RELATED REGULATIONS

Developments of wireless technology for medical devices is elevating the provision of healthcare with lower expenses. The wireless telecommunications can be used for both wearable and implantable applications such as deep brain stimulation (DBS), tracking of vital signs, measuring biological parameters, and cardiac rhythm control. The main advantage of wireless technology compared to landline networks is that the patient is not required to be linked to a certain location by cables [110]. Despite advances in biomedical implants such as a pacemaker, cochlear implant, and nerve stimulator these devices need to be improved in terms of miniaturization, the biocompatibility of materials, sources of electric charge, and wireless communication. To develop an effective implantable medical device, the doctor, the patient, and the technician must collaborate in collecting coherent initial information about different aspects of the device. In particular, the user's satisfaction, doctor's technical priorities, and the workability of the model are necessary to be considered in the design process [111]. There are important factors for designing medical implants. Since an electric device is implanted inside the human body, the organisms around the device may react to it. To avoid such an issue, the device should be made up of biocompatible materials. Moreover, the medical implants should have appropriate packaging to isolate components of the device from body tissue. Another factor is the structure of the design itself. Before the design, enough data should be collected from the patients, engineers, previous designs and their advantages and drawbacks [111]. The United States' medical devices market is regulated by the three different organizations, Federal Medicines Authorities (FCC), the Food and Drug Administration (FDA) and the Centers for Medicare and Medicaid Services (CMS). Wireless medical instruments can be classified into two categories, short-range such as inductive implants, medical body area networks and long-range such as wireless medical telemetry (WMTS). According to the Federal Communications Commission (FCC), short-range technology sends data to the local receivers and long-range technology sends user data to the remote spot [110]. The FDA's mission is to check if the proposed medical devices guarantee the factors of safeness and effectiveness for patient usage. The FDA divides medical devices into three classifications based on the risk factor. Class I includes the lowest-risk devices and without FDA prior authorization, medical devices may be advertised in this class. The medical devices using wireless technologies usually considered in Class II. The highest-risk medical devices fall under Class III and clinical trials are mandatory to get FDA approval. The FCC and FDA must permit before wireless medical devices can be marketed in the United States. It is worth mentioning that the FDA and FCC have distinct criteria, and one agency's authorization does not simply ensure the other's consent [110]. The designers of medical implants currently dealing with challenges in materials, output power, size miniaturization, and efficiency of wireless link [111]. It is crucial to have a broad perspective of different aspects of wireless techniques before choosing the tactic for any specific applications. Figure 9 shows a comparison of different approaches based on the maximum dimension of a receiver, power transfer efficiency, and frequency. It can be interpreted that the maximum power transmission efficiency is achieved via inductive-based wireless power transfer. Besides, a cluster of biomedical implants' receivers using the inductive-based technique have a maximum dimension of fewer than 20 millimeters operating at a quite lower frequency compared to microwave-based technique with the almost same size.

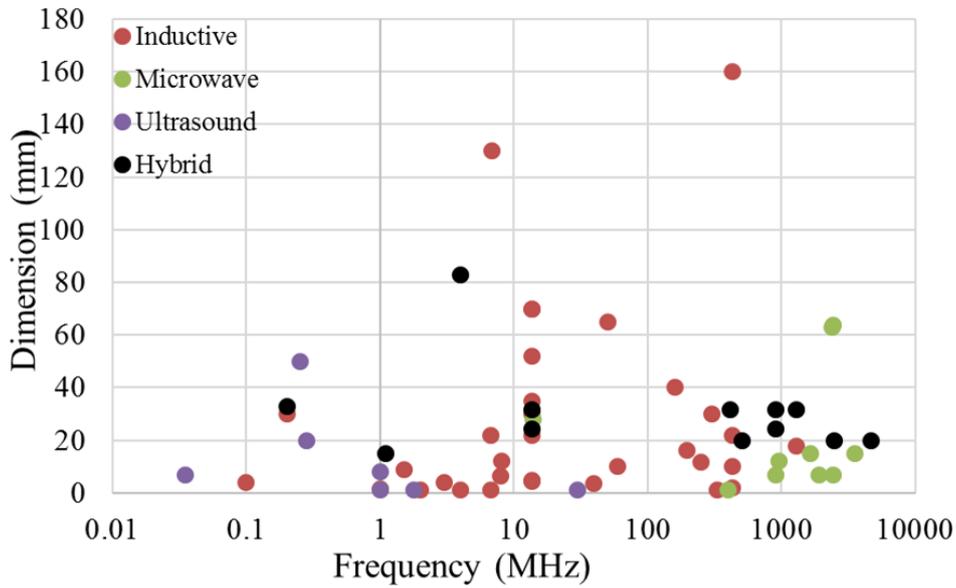

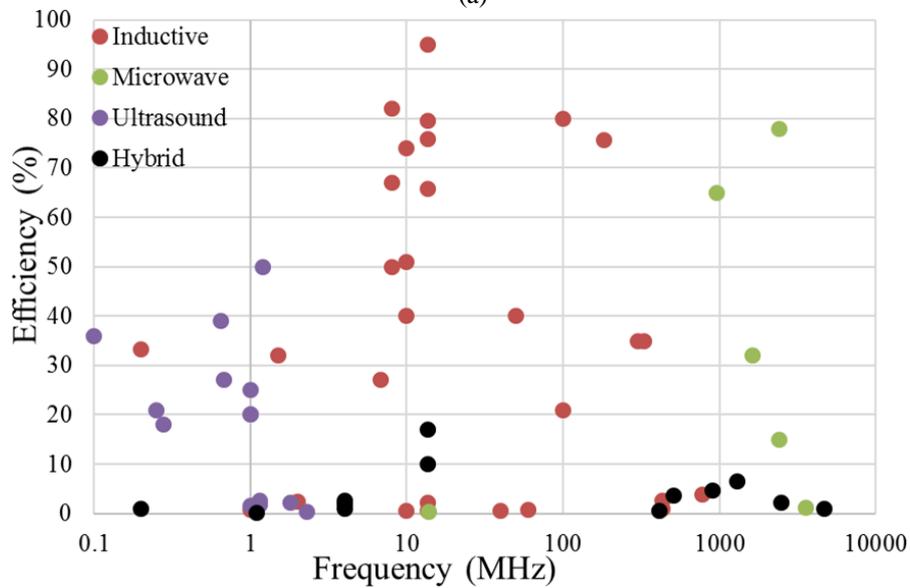

Figure 9: A comparisons of different approaches. (a) Maximum dimension of receiver versus frequency (b) efficiency versus frequency.

## 4. CONCLUSION

This research has evaluated and discussed a survey of popular methods for wirelessly transfer power into implantable medical devices. Power delivered in the reviewed works to medical implants varies from few μW to 5.4 W with a distance of 1 mm to 4 m and maximum efficiency of up to 79.6%. This paper provides a perspective on different WPT approaches for biomedical applications by investigating the significant works in this field. Also, some points for designing an effective implantable medical device and related commercial rules and regulations are presented.

## References

[1] Amar, A. Ben; Kouki, A.B.; Cao, H. Power approaches for implantable medical devices. Sensors (Switzerland) 2015, 15, 28889–28914.

[2] R. V. Taalla, M. S. Arefin, A. Kaynak and A. Z. Kouzani, "A Review on Miniaturized Ultrasonic Wireless Power Transfer to Implantable Medical Devices," in IEEE Access, vol. 7, pp. 2092-2106, 2019.

[3] Shadid, R.; Noghanian, S. A Literature Survey on Wireless Power Transfer for Biomedical Devices. Int. J. Antennas Propag. 2018.

[4] S. Mutashar, M. Hannan, S. Samad, and A. Hussain, "Analysis and optimization of spiral circular inductive coupling link for bio-implanted applications on air and within human tissue," Sensors, vol. 14, no. 7, pp. 11522–11541, 2014.

[5] Y. Yu, H. Hao, W. Wang, and L. Li, ``Simulative and experimental research on wireless power transmission technique in implantable medical device,'' presented at the IEEE Annu. Int. Conf. Eng. Med. Biol. Soc., Minneapolis, MN, USA, Sep. 2009.

[6] Bocan, K.N.; Sejdić, E. Adaptive Transcutaneous Power Transfer to Implantable Devices: A State of the Art Review. Sensors 2016, 16, 393.

[7] H. Lyu, P. Gad, H. Zhong, V. R. Edgerton and A. Babakhani, "A 430-MHz Wirelessly Powered Implantable Pulse Generator With Intensity/Rate Control and Sub-1 µA Quiescent Current Consumption," in IEEE Transactions on Biomedical Circuits and Systems, vol. 13, no. 1, pp. 180-190, Feb. 2019.

[8] H. Lyu, J. Wang, J. La, J. M. Chung and A. Babakhani, "An Energy-Efficient Wirelessly Powered Millimeter-Scale Neurostimulator Implant Based on Systematic Codesign of an Inductive Loop Antenna and a Custom Rectifier," in IEEE Transactions on Biomedical Circuits and Systems, vol. 12, no. 5, pp. 1131-1143, Oct. 2018. doi: 10.1109/TBCAS.2018.2852680

[9] A. Khalifa et al., "The Microbead: A Highly Miniaturized Wirelessly Powered Implantable Neural Stimulating System," in IEEE Transactions on Biomedical Circuits and Systems, vol. 12, no. 3, pp. 521-531, June 2018.

[10] S. A. Mirbozorgi, P. Yeon and M. Ghovanloo, "Robust Wireless Power Transmission to mm-Sized Free-Floating Distributed Implants," in IEEE Transactions on Biomedical Circuits and Systems, vol. 11, no. 3, pp. 692-702, June 2017. doi: 10.1109/TBCAS.2017.2663358

[11] C. Yang, C. Chang, S. Lee, S. Chang and L. Chiou, "Efficient Four-Coil Wireless Power Transfer for Deep Brain Stimulation," in IEEE Transactions on Microwave Theory and Techniques, vol. 65, no. 7, pp. 2496-2507, July 2017.

[12] P. J. Larson and B. C. Towe, ``Miniature ultrasonically powered wireless nerve cuff stimulator,'' in Proc. 5th IEEE/EMBS Int. Conf. Neural Eng. (NER), Cancun, Mexico, Apr./May 2011, pp. 265 268.

[13] S. M. Asif, A. Iftikhar, J. W. Hansen, M. S. Khan, D. L. Ewert and B. D. Braaten, "A Novel RF-Powered Wireless Pacing via a Rectenna-Based Pacemaker and a Wearable Transmit-Antenna Array," in IEEE Access, vol. 7, pp. 1139-1148, 2019. doi: 10.1109/ACCESS.2018.2885620

[14] M. Zargham and P. G. Gulak, "A 0.13µm CMOS integrated wireless power receiver for biomedical applications," in 2013 Proceedings of the ESSCIRC (ESSCIRC), pp. 137– 140, Bucharest, Romania, 2013.

[15] K. Keikhosravy, P. Kamalinejad, S. Mirabbasi, K. Takahata, and V. C. M. Leung, ``An ultra-low-power monitoring system for inductively coupled biomedical implants,'' presented at the IEEE Int. Symp. Circuits Syst. (ISCAS), Beijing, China, 2013.

[16] F. Mazzilli and C. Dehollain, ``184 W ultrasonic on off keying/amplitude-shift keying demodulator for downlink communication in deep implanted medical devices,'' Electron. Lett., vol. 52, no. 7, pp. 502 504, 2016.

[17] J. Charthad, M. J. Weber, T. C. Chang, and A. Arbabian, ``A mm-sized implantable medical device (IMD) with ultrasonic power transfer and a hybrid bi-directional data link,'' IEEE J. Solid-State Circuits, vol. 50, no. 8, pp. 1741 1753, Aug. 2015.

[18] F. Mazzilli, C. Lafon, and C. Dehollain, ``A 10.5 cm ultrasound link for deep implanted medical devices,'' IEEE Trans. Biomed. Circuits Syst., vol. 8, no. 5, pp. 738 750, Oct. 2014.

[19] D. Ahn and S. Hong, "Wireless power transmission with self-regulated output voltage for biomedical implant," IEEE Transactions on Industrial Electronics, vol. 61, no. 5, pp. 2225–2235, 2014.

[20] E. G. Kilinc, M. A. Ghanad, F. Maloberti, and C. Dehollain, ``Short range remote powering of implanted electronics for freely moving animals,'' presented at the IEEE 11th Int. New Circuits Syst. Conf. (NEWCAS), Paris, France, Jun. 2013.

[21] H. Lyu, Z. Jian, X. Liu, Y. Sun and A. Babakhani, "Towards the Implementation of a Wirelessly Powered Dielectric Sensor With Digitized Output for Implantable Applications," in IEEE Sensors Letters, vol. 3, no. 3, pp. 1-4, March 2019, Art no. 5500204. doi: 10.1109/LSENS.2019.2896185

[22] E. G. Kilinc, C. Dehollain, and F. Maloberti, ``A low-power PPM demodulator for remotely powered batteryless implantable devices,'' presented at the IEEE 57th Int. Midwest Symp. Circuits Syst. (MWSCAS), College Station, TX, USA, Aug. 2014.

[23] M. Zargham and P. G. Gulak, ``Fully integrated on-chip coil in 0.13 µm CMOS for wireless power transfer through biological media,'' IEEE Trans. Biomed. Circuits Syst., vol. 9, no. 2, pp. 259_271, Apr. 2015.

[24] M. A. Adeeb, A. B. Islam, M. R. Haider, F. S. Tulip, M. N. Ericson, and S. K. Islam, ``An inductive link-based wireless power transfer system for biomedical applications,'' Act. Passive Electron. Compon., vol. 2012, Mar. 2012, Art. no. 879294, doi: 10.1155/2012/879294.

[25] S.-N. Suzuki, S. Kimura, T. Katane, H. Saotome, O. Saito, and K. Kobayashi, ``Power and interactive information transmission to implanted medical device using ultrasonic,'' Jpn. J. Appl. Phys., vol. 41, no. 5B, Jan. 2002.

[26] H. Kawanabe, T. Katane, H. Saotome, O. Saito, and K. Kobayashi, ``Power and information transmission to implanted medical device using ultrasonic,'' Jpn. J. Appl. Phys., vol. 40, no. 5B, p. 3865, 2001.

[27] E. G. Kilinc et al., ``Remotely powered implantable heart monitoring system for freely moving animals,'' presented at the IEEE 5th Int. Workshop Adv. Sensors Interfaces. (IWASI), Bari, Italy, 2013.

[28] G. Simard, M. Sawan, and D. Massicotte, ``High-speed OQPSK and ef cient power transfer through inductive link for biomedical implants,''IEEE Trans. Biomed. Circuits Syst., vol. 4, no. 3, pp. 192 200, Jun. 2010.

[29] D. Jiang, D. Cirmirakis, M. Schormans, T. A. Perkins, N. Donaldson, and A. Demosthenous, ``An integrated passive phase-shift keying modulator for biomedical implants with power telemetry over a single inductive link," IEEE Trans. Biomed. Circuits Syst., vol. 11, no. 1, pp. 64_77, Feb. 2017.

[30] M. Kiani, B. Lee, P. Yeon, and M. Ghovanloo, ``A Q-modulation technique for ef_cient inductive power transmission,'' IEEE J. Solid-State Circuits, vol. 50, no. 12, pp. 2839_2848, Dec. 2015.

[31] Z. Lu and M. Sawan, ``An 8 Mbps data rate transmission by inductive link dedicated to implantable devices,'' presented at the IEEE Int. Symp. Circuits Syst., Seattle, WA, USA, May 2008.

[32] S. Mandal and R. Sarpeshkar, ``Power-ef cient impedance-modulation wireless data links for biomedical implants,'' IEEE Trans. Biomed. Circuits Syst., vol. 2, no. 4, pp. 301 315, Dec. 2008.

[33] H. Rahmani and A. Babakhani, "A Dual-Mode RF Power Harvesting System With an On-Chip Coil in 180-nm SOI CMOS for Millimeter-Sized Biomedical Implants," in IEEE Transactions on Microwave Theory and Techniques, vol. 67, no. 1, pp. 414-428, Jan. 2019.

[34] G. Sun, B. Muneer, Y. Li and Q. Zhu, "Ultracompact Implantable Design With Integrated Wireless Power Transfer and RF Transmission Capabilities," in IEEE Transactions on Biomedical Circuits and Systems, vol. 12, no. 2, pp. 281-291, April 2018. doi: 10.1109/TBCAS.2017.2787649

[35] Institute of Electrical and Electronics Engineers IEEE Standard for Safety Levels with Respect to Human Exposure to Radio Frequency Electromagnetic Fields, 3 kHz to 300 GHz.; 1999; ISBN 155937179X

[36] A. Ibrahim and M. Kiani, ``A figure-of-merit for design and optimization of inductive power transmission links for millimeter-sized biomedical implants,'' IEEE Trans. Biomed. Circuits Syst., vol. 10, no. 6, pp. 1100_1111, Dec. 2016.

[37] M. Manoufali, K. Bialkowski, B. Mohammed and A. Abbosh, "Wireless Power Link Based on Inductive Coupling for Brain Implantable Medical Devices," in IEEE Antennas and Wireless Propagation Letters, vol. 17, no. 1, pp. 160-163, Jan. 2018. doi: 10.1109/LAWP.2017.2778698

[38] P. Chen, H. Yang, R. Luo and B. Zhao, "A Tissue-Channel Transcutaneous Power Transfer Technique for Implantable Devices," in IEEE Transactions on Power Electronics, vol. 33, no. 11, pp. 9753-9761, Nov. 2018.

[39] A. Ibrahim and M. Kiani, ``Inductive power transmission to millimeter-sized biomedical implants using printed spiral coils,'' presented at the IEEE 38th Annu. Int. Conf. Eng. Med. Biol. Soc. (EMBC), Orlando, FL, USA, Aug. 2016.

[40] P. Feng, P. Yeon, Y. Cheng, M. Ghovanloo and T. G. Constandinou, "Chip-Scale Coils for Millimeter-Sized Bio-Implants," in IEEE Transactions on Biomedical Circuits and Systems, vol. 12, no. 5, pp. 1088-1099, Oct. 2018.

[41] J. Faerber et al., "In Vivo Characterization of a Wireless Telemetry Module for a Capsule Endoscopy System Utilizing a Conformal Antenna," in IEEE Transactions on Biomedical Circuits and Systems, vol. 12, no. 1, pp. 95-105, Feb. 2018. doi: 10.1109/TBCAS.2017.2759254

[42] H. S. Gougheri and M. Kiani, ``Optimal frequency for powering millimeter-sized biomedical implants inside an inductively-powered homecage,'' presented at the IEEE 38th Annu. Int. Conf. Eng. Med. Biol. Soc., Aug. 2016.

[43] K. M. Silay, C. Dehollain, and M. Declercq, ``A closed-loop remote powering link for wireless cortical implants,'' IEEE Sensors J., vol. 13, no. 9, pp. 3226 3235, Sep. 2013.

[44] B. M. G. Rosa and G. Z. Yang, ``Active implantable sensor powered by ultrasounds with application in the monitoring of physiological parameters for soft tissues,'' presented at the IEEE 13th Int. Conf. Wearable Implant. Body Sensor Netw. (BSN), San Francisco, CA, USA, Jun. 2016.

[45] T. Maleki, N. Cao, S. H. Song, C. Kao, S.-C. Ko, and B. Ziaie, ``An ultrasonically powered implantable micro-oxygen generator (IMOG),'' IEEE Trans. Biomed. Eng., vol. 58, no. 11, pp. 3104 3111, Nov. 2011.

[46] Shadid, R.; Haerinia, M.; Noghanian, S. Study of Rotation and Bending Effects on a Flexible Hybrid Implanted Power Transfer and Wireless Antenna System. Sensors 2020, 20, 1368. doi: 10.3390/s20051368

[47] K. N. Bocan, M. H. Mickle and E. Sejdić, "Multi-Disciplinary Challenges in Tissue Modeling for Wireless Electromagnetic Powering: A Review," in IEEE Sensors Journal, vol. 17, no. 20, pp. 6498-6509, 15Oct.15,2017.

[48] C. Yang, W. Li and R. K. Chen, "Determination of Tissue Thermal Conductivity as a Function of Thermal Dose and Its Application in Finite Element Modeling of Electrosurgical Vessel Sealing," in IEEE Transactions on Biomedical Engineering. doi: 10.1109/TBME.2020.2972465

[49] Hao, J.J., Lv, L.J., Ju, L., Xie, X., Liu, Y.J., and Yang, H.W. (2019). Simulation of microwave propagation properties in human abdominal tissues on wireless capsule endoscopy by FDTD. Biomedical Signal Processing and Control 49, 388–395.

[50] A. Fornes-Leal et al., "Dielectric Characterization of In Vivo Abdominal and Thoracic Tissues in the 0.5–26.5 GHz Frequency Band for Wireless Body Area Networks," in IEEE Access, vol. 7, pp. 31854-31864, 2019.

[51] J. Braun, H. Tzschatzsch, C. Korting, A. A. de Schellenberger, M. Jenderka, T. Driessle, M. Ledwig, and I. Sack, "A compact 0.5 T MR elastography device and its application for studying viscoelasticity changes in biological tissues during progressive formalin fixation," Magn. Reson. Med. 79, 470–478 (2018). doi:10.1002/mrm.26659

[52] K. N. Bocan, M. H. Mickle and E. Sejdić, "Simulating, Modeling, and Sensing Variable Tissues for Wireless Implantable Medical Devices," in IEEE Transactions on Microwave Theory and Techniques, vol. 66, no. 7, pp. 3547-3556, July 2018. doi: 10.1109/TMTT.2018.2811497

[53] Li Gun, Du Ning, and Zhang Liang, "Effective Permittivity of Biological Tissue: Comparison of Theoretical Model and Experiment," Mathematical Problems in Engineering, vol. 2017, Article ID 7249672, 7 pages, 2017.

[54] E. Balidemaj et al., "In vivo electric conductivity of cervical cancer patients based on B+1 maps at 3T MRI," Phys. Med. Biol., vol. 61, no. 4, p. 1596, 2016.

[55] S. Dahdouh, N. Varsier, M. A. N. Ochoa, J. Wiart, A. Peyman, and I. Bloch, "Infants and young children modeling method for numerical dosimetry studies: Application to plane wave exposure," Phys. Med. Biol., vol. 61, no. 4, p. 1500, 2016.

[56] A. T. Mobashsher, A. Mahmoud, and A. M. Abbosh, "Portable wideband microwave imaging system for intracranial hemorrhage detection using improved back-projection algorithm with model of effective head permittivity," Sci. Rep., vol. 6, Feb. 2016, Art. no. 20459.

[57] M. T. Jilani, W. P. Wen, L. Y. Cheong, and M. Z. ur Rehman, "A microwave ring-resonator sensor for non-invasive assessment of meat aging," Sensors, vol. 16, no. 1, p. 52, 2016.

[58] Y. Diao, S.-W. Leung, K. H. Chan, W. Sun, Y.-M. Siu, and R. Kong, "The effect of gaze angle on the evaluations of SAR and temperature rise in human eye under plane-wave exposures from 0.9 to 10 GHz," Radiat. Protection Dosimetry, vol. 172, no. 4, pp. 393–400, 2015.

[59] A. F. Abdelaziz, Q. H. Abbasi, K. Yang, K. Qaraqe, Y. Hao, and A. Alomainy, "Terahertz signal propagation analysis inside the human skin," in Proc. IEEE 11th Int. Conf. Wireless Mobile Comput., Netw. Commun. (WiMob), Oct. 2015, pp. 15–19.

[60] C. Li, Q. Chen, Y. Xie, and T. Wu, "Dosimetric study on eye's exposure to wide band radio frequency electromagnetic fields: Variability by the ocular axial length," Bioelectromagnetics, vol. 35, no. 5, pp. 324–336, 2014.

[61] X. Zhang, S. Zhu and B. He, "Imaging Electric Properties of Biological Tissues by RF Field Mapping in MRI," in IEEE Transactions on Medical Imaging, vol. 29, no. 2, pp. 474-481, Feb. 2010.

[62] A. Peyman, C. Gabriel, E. H. Grant, G. Vermeeren, and L. Martens, "Variation of the dielectric properties of tissues with age: The effect on the values of SAR in children when exposed to walkieâŁ"talkie devices," Phys. Med. Biol., vol. 54, no. 2, pp. 227–241, 2009.

[63] J. Keshvari, R. Keshvari, and S. Lang, "The effect of increase in dielectric values on specific absorption rate (SAR) in eye and head tissues following 900, 1800 and 2450 MHz radio frequency (RF)

exposure," Phys. Med. Biol., vol. 51, no. 6, p. 1463, 2006.

[64] J. Ryckaert, P. D. Doncker, R. Meys, A. D. L. Hoye, and S. Donnay, "Channel model for wireless communication around human body," Electron. Lett., vol. 40, no. 9, pp. 543–544, Apr. 2004.

[65] T. Nagaoka et al., "Development of realistic high-resolution whole-body voxel models of Japanese adult males and females of average height and weight, and application of models to radio-frequency electromagneticfield dosimetry," Phys. Med. Biol., vol. 49, no. 1, pp. 1–15, 2004.

[66] V. Monebhurrun, C. Dale, J. C. Bolomey, and J. Wiart, "A numerical approach for the determination of the tissue equivalent liquid used during SAR assessments," IEEE Trans. Magn., vol. 38, no. 2, pp. 745–748, Mar. 2002.

[67] H. Ali, T. J. Ahmad, and S. A. Khan, ``Inductive link design for medical implants,'' presented at the IEEE Symp. Ind. Electron. Appl., Kuala Lumpur, Malaysia, Oct. 2009.

[68] Z. N. Low, R. A. Chinga, R. Tseng, and J. Lin, ``Design and test of a high power high-ef ciency loosely coupled planar wireless power transfer system,'' IEEE Trans. Ind. Electron., vol. 56, no. 5, pp. 1801 1812, May 2009.

[69] K. Zhang et al., "Near-Field Wireless Power Transfer to Deep-Tissue Implants for Biomedical Applications," in IEEE Transactions on Antennas and Propagation, vol. 68, no. 2, pp. 1098-1106, Feb. 2020.

doi: 10.1109/TAP.2019.2943424

[70] B. Zhao, N. Kuo, B. Liu, Y. Li, L. Iotti and A. M. Niknejad, "A Batteryless Padless Crystalless 116µ ×116 µm "Dielet" Near-Field Radio With On-Chip Coil Antenna," in IEEE Journal of Solid-State Circuits, vol. 55, no. 2, pp. 249-260, Feb. 2020. doi: 10.1109/JSSC.2019.2954772

[71] L. L. Pon, S. K. Abdul rahim, C. Y. Leow, M. Himdi and M. Khalily, "Displacement-Tolerant Printed Spiral Resonator With Capacitive Compensated-Plates for Non-Radiative Wireless Energy Transfer," in IEEE Access, vol. 7, pp. 10037-10044, 2019. doi: 10.1109/ACCESS.2019.2891015

[72] M. M. Ahmadi and S. Ghandi, "A Class-E Power Amplifier With Wideband FSK Modulation for Inductive Power and Data Transmission to Medical Implants," in IEEE Sensors Journal, vol. 18, no. 17, pp. 7242-7252, 1 Sept.1, 2018. doi: 10.1109/JSEN.2018.2851605

[73] N. A. Quadir, L. Albasha, M. Taghadosi, N. Qaddoumi and B. Hatahet, "Low-Power Implanted Sensor for Orthodontic Bond Failure Diagnosis and Detection," in IEEE Sensors Journal, vol. 18, no. 7, pp. 3003-3009, 1 April1, 2018. doi: 10.1109/JSEN.2018.2791426

[74] L. Li, H. Liu, H. Zhang and W. Xue, "Efficient Wireless Power Transfer System Integrating With Metasurface for Biological Applications," in IEEE Transactions on Industrial Electronics, vol. 65, no. 4, pp. 3230-3239, April 2018. doi: 10.1109/TIE.2017.2756580

[75] Haerinia, M.; Afjei, E.S. Resonant inductive coupling as a potential means for wireless power transfer to printed spiral coil. J. Electr. Eng. 2016, 16, 65–74.

[76] Haerinia, M. Modeling and simulation of inductive-based wireless power transmission systems. In Book Energy Harvesting for Wireless Sensor Networks: Technology, Components and System Design, 1st ed.; Olfa, K., Ed.; De Gruyter: Berlin, Germany; Boston, MA, USA, 2018; pp. 197–220.

[77] Haerinia, M.; Afjei, E.S. Design and analysis of class EF2 inverter for driving transmitting printed spiral coil. J. Electr. Eng. 2018, 18, 1–5.

[78] Y. Cheng, G. Chen, D. Xuan, G. Qian, M. Ghovanloo and G. Wang, "Analytical Modeling of Small, Solenoidal, and Implantable Coils With Ferrite Tube Core," in IEEE Microwave and Wireless Components Letters, vol. 29, no. 3, pp. 237-239, March 2019. doi: 10.1109/LMWC.2019.2891620

[79] Haerinia, M.; Mosallanejad, A.; Afjei, E.S. Electromagnetic analysis of different geometry of transmitting coils for wireless power transmission applications. Prog. Electromagn. Res. M 2016, 50, 161–168.

[80] X. Wang and M. Lu, "Microwave Power Transmission Based on Retro-reflective Beamforming," Wireless Power Transfer—Fundamentals and Technologies; Coca, E., Ed.; In Tech: London, UK, 2016.

[81] M. Zada and H. Yoo, "A Miniaturized Triple-Band Implantable Antenna System for Bio-Telemetry Applications," in IEEE Transactions on Antennas and Propagation, vol. 66, no. 12, pp. 7378-7382, Dec. 2018.

[82] Asif, S.M.; Iftikhar, A.; Braaten, B.D.; Ewert, D.L.; Maile, K. A Wide-Band Tissue Numerical Model for Deeply Implantable Antennas for RF-Powered Leadless Pacemakers. IEEE Access 2019, 7, 1–1.

[83] Patlolla, B.; Yeh, A.J.; Beygui, R.E.; Poon, A.S.Y.; Tanabe, Y.; Neofytou, E.; Kim, S.; Ho, J.S. Wireless power transfer to deep-tissue microimplants. Proc. Natl. Acad. Sci. 2014, 111, 7974–7979.

[84] A. Basir and H. Yoo, "Efficient Wireless Power Transfer System With a Miniaturized Quad-Band Implantable Antenna for Deep-Body Multitasking Implants," in IEEE Transactions on Microwave Theory and Techniques.
doi: 10.1109/TMTT.2020.2965938

[85] Y. Fan, H. Liu, X. Liu, Y. Cao, Z. Li and M. M. Tentzeris, "Novel coated differentially fed dual-band fractal antenna for implantable medical devices," in IET Microwaves, Antennas & Propagation, vol. 14, no. 2, pp. 199-208, 5 2 2020. doi: 10.1049/iet-map.2018.6171

[86] Haerinia, M.; Noghanian, S. Analysis of misalignment effects on link budget of an implantable antenna. In Proceedings of the URSI EM Theory Symposium, EMTS 2019, San Diego, CA, USA, 27–31 May 2019. Accepted

[87] Haerinia, Mohammad, and Sima Noghanian. "A Printed Wearable Dual-Band Antenna for Wireless Power Transfer." Sensors (Basel, Switzerland) vol. 19,7 1732. 11 Apr. 2019, doi: 10.3390/s19071732

[88] Haerinia, M.; Noghanian, S. Study of Bending Effects on a Dual-Band Implantable Antenna. 2019 IEEE International Symposium on Antennas and Propagation and USNC-URSI Radio Science Meeting, Atlanta, Georgia, USA, 09 July 2019.

[89] A. Aldaoud et al., "A Stent-Based Power and Data Link for Sensing Intravascular Biological Indicators," in IEEE Sensors Letters, vol. 2, no. 4, pp. 1-4, Dec. 2018, Art no. 7501004. doi: 10.1109/LSENS.2018.2876350

[90] S. Q. Lee, W. Youm, and G. Hwang, ``Biocompatible wireless power transferring based on ultrasonic resonance devices,'' in Proc. Meetings Acoust., vol. 19, no. 1, 2013, p. 030030.

[91] A. Kim, C. R. Powell, and B. Ziaie, ``An implantable pressure sensing system with electromechanical interrogation scheme,'' IEEE Trans. Biomed. Eng., vol. 61, no. 7, pp. 2209 2217, Jul. 2014.

[92] S. H. Song, A. Kim, and B. Ziaie, ``Omnidirectional ultrasonic powering for millimeter-scale implantable devices,'' IEEE Trans. Biomed. Eng., vol. 62, no. 11, pp. 2717 2723, Nov. 2015.

[93] G. E. Santagati, N. Dave and T. Melodia, "Design and Performance Evaluation of an Implantable Ultrasonic Networking Platform for the Internet of Medical Things," in IEEE/ACM Transactions on Networking, vol. 28, no. 1, pp. 29-42, Feb. 2020. doi: 10.1109/TNET.2019.2949805

[94] T. C. Chang, M. L. Wang, J. Charthad, M. J. Weber, and A. Arbabian, ``A 30.5 mm$^3$ fully packaged implantable device with duplex ultrasonic data and power links achieving 95 kb/s with <$10^{-4}$BER at 8.5 cm depth,''

in IEEE Int. Solid-State Circuits Conf. (ISSCC) Dig. Tech. Papers, San Francisco, CA, USA, Feb. 2017, pp. 460 461.

[95] M. Meng and M. Kiani, ``Design and optimization of ultrasonic wireless power transmission links for millimeter-sized biomedical implants,'' IEEE Trans. Biomed. Circuit Syst., vol. 11, no. 1, pp. 98 107, Feb. 2017.

[96] H. Vihvelin, J. R. Leadbetter, M. Bance, J. A. Brown, and R. B. A. Adamson, ``Compensating for tissue changes in an ultrasonic power link for implanted medical devices,'' IEEE Trans. Biomed. Circuits

Syst., vol. 10, no. 2, pp. 404 411, Apr. 2016.

[97] S. Q. Lee, W. Youm, G. Hwang, and K. S. Moon, ``Wireless power transferring and charging for implantable medical devices based on ultrasonic resonance,'' in Proc. 22nd Int. Congr. Sound Viberation, 2015, pp. 1 7.

[98] B. Fang, T. Feng, M. Zhang, and S. Chakrabartty, ``Feasibility of B-mode diagnostic ultrasonic energy transfer and telemetry to a cm2 sized deep-tissue implant,'' presented at the IEEE Int. Symp. Circuits Syst. (ISCAS), Lisbon, Portugal, May 2015.

[99] J. Leadbetter, J. Brown, and R. Adamson, ``The design of ultrasonic lead magnesium niobate-lead titanate composite transducers for power and signal delivery to implanted hearing aids,'' J. Acoust. Soc. Amer., vol. 133, no. 5, p. 3268, May 2013.

[100] Y. Shigeta, Y. Hori, K. Fujimori, K. Tsuruta, and S. Nogi, ``Development of highly efficient transducer for wireless power transmission system by ultrasonic,'' presented at the IEEE MTT-S Int. Microw. Workshop Ser. Innov.Wireless Power Transmiss., Technol., Syst., Appl., Kyoto, Japan, May 2011.

[101] P.-J. Shih and W.-P. Shih, ``Design, fabrication, and application of bioimplantable acoustic power transmission,'' J. Microelectromech. Syst., vol. 19, no. 3, pp. 494 502, Jun. 2010.

[102] S. Ozeri and D. Shmilovitz, ``Ultrasonic transcutaneous energy transfer for powering implanted devices,'' Ultrasonics, vol. 50, no. 6, pp. 556 566, May 2010.

[103] S. N. Suzuki, T. Katane, and O. Saito, ``Fundamental study of an electric power transmission system for implanted medical devices using magnetic and ultrasonic energy,'' J. Artif. Organs, vol. 6, no. 2, pp. 145 148, 2003.

[104] Haerinia, M.; Noghanian, S. Design of hybrid wireless power transfer and dual ultrahigh-frequency antenna system. In Proceedings of the URSI EM Theory Symposium, EMTS 2019, San Diego, CA, USA, 27–31 May 2019.

[105] M. Meng and M. Kiani, "A Hybrid Inductive-Ultrasonic Link for Wireless Power Transmission to Millimeter-Sized Biomedical Implants," in IEEE Transactions on Circuits and Systems II: Express Briefs, vol. 64, no. 10, pp. 1137-1141, Oct. 2017. doi: 10.1109/TCSII.2016.2626151

[106] A. Aldaoud et al., "Near-Field Wireless Power Transfer to Stent-Based Biomedical Implants," in IEEE Journal of Electromagnetics, RF and Microwaves in Medicine and Biology, vol. 2, no. 3, pp. 193-200, Sept. 2018.

[107] Shadid, R.; Haerinia, M.; Sayan, R.; Noghanian, S. Hybrid Inductive Power Transfer and Wireless Antenna System for Biomedical Implanted Devices. Prog. Electromagn. Res. C 2018, 88, 77–88.

[108] A. Sharma, E. Kampianakis, and M. S. Reynolds, "A dual-band HF and UHF antenna system for implanted neural recording and stimulation devices," IEEE Antennas and Wireless Propagation Letters, vol. 16, pp. 493–496, 2017.

[109] A. Sanni, A. Vilches, and C. Toumazou, ``Inductive and ultrasonic multi-tier interface for low-power, deeply implantable medical devices,'' IEEE Trans. Biomed. Circuits Syst., vol. 6, no. 4, pp. 297 308, Aug. 2012.

[110] 'Fish & Richardson' [Online]. Available: www.fr.com

[111] Joung YH. Development of implantable medical devices: from an engineering perspective. Int Neurourol J Sep 2013;17(3):98–106.

[112] K. Li, K. See, W. Koh and J. Zhang, "Design of 2.45 GHz microwave wireless power transfer system for battery charging applications," 2017 Progress in Electromagnetics Research Symposium - Fall (PIERS - FALL), Singapore, 2017, pp. 2417-2423. doi: 10.1109/PIERS-FALL.2017.8293542